# The role of preprints in open science: Accelerating knowledge transfer from science to technology

Zhiqi Wang[a,b,*], Yue Chen[b,*], Chun Yang[b]

[a] School of Management, Liaoning Normal University, Dalian, 116081, China

[b] Institute of Science of Science and S&T Management & WISELab, Dalian University of Technology, Dalian, 116024, China

**Abstract**
Preprints have become increasingly essential in the landscape of open science, facilitating not only the exchange of knowledge within the scientific community but also bridging the gap between science and technology. However, the impact of preprints on technological innovation, given their unreviewed nature, remains unclear. This study fills this gap by conducting a comprehensive scientometric analysis of patent citations to bioRxiv preprints submitted between 2013 and 2021, measuring and accessing the contribution of preprints in accelerating knowledge transfer from science to technology. Our findings reveal a growing trend of patent citations to bioRxiv preprints, with a notable surge in 2020, primarily driven by the COVID-19 pandemic. Preprints play a critical role in accelerating innovation, not only expedite the dissemination of scientific knowledge into technological innovation but also enhance the visibility of early research results in the patenting process, while journals remain essential for academic rigor and reliability. The substantial number of post-online-publication patent citations highlights the critical role of the open science model—particularly the "open access" effect of preprints—in amplifying the impact of science on technological innovation. This study provides empirical evidence that open science policies encouraging the early sharing of research outputs, such as preprints, contribute to more efficient linkage between science and technology, suggesting an acceleration in the pace of innovation, higher innovation quality, and economic benefits.

\* Corresponding author.

E-mail addresses: zhiqi_wang90@126.com (Z. Wang), chenyuedlut@163.com (Y. Chen).

# 1. Introduction

Preprints are integral components of the open science ecosystem, designed to rapidly and efficiently disseminate the lasted research findings to relevant audiences. Historically, preprints have played a crucial role in fields, such as Physics, Mathematics and Computer Science (Ginsparg, 1994, 2011). More recently, in the life sciences, the bioRxiv preprint server was launched in 2013, followed by medRxiv preprint server, both of which saw a significant increase in usage during the COVID-19 pandemic (Fraser et al., 2022). There is substantial evidence supporting the vital role of preprints in scholarly communication, including their ability to accelerate knowledge dissemination, broaden the reach and impact of research, and promote academic autonomy (Fraser et al., 2020; Wang, 2019; Wang, Chen, et al., 2020; Wang, Glänzel, et al., 2020). In the new era of open science, open access to scientific knowledge is recognized for its potential to broaden access to information, benefit both science and society, and foster innovation and participation in the co-creation of knowledge (United Nations Educational, 2021). This context highlights the social impact of research output, especially the contribution of scientific knowledge to today's innovation.

Exploring the connection between science and technology enhances our understanding of how the flow and integration of knowledge drive innovation. Whereas citations between research papers capture knowledge flow within the academic community and are commonly used to assess the research impact of scientific publications (Glänzel & Chi, 2016; Glänzel & Schoepflin, 1995), patent citations to scientific literature – referred to as scientific non-patent references (SNPRs) – more directly captures the knowledge transfer from basic scientific research to technological innovation, a critical precursor to commercial products (Cao et al., 2023; Higham et al., 2021; Li et al., 2014; Munari et al., 2022; Yin et al., 2022). Therefore, SNPRs provide a complementary measure of the broader impact of academic research beyond traditional scientific communication. This is particularly important because some scholarly publications hold commercial value or have real-world applications that do not directly translate into academic citations.

Our research contributes to the literature as the first study to explore patents citing preprints, providing new insights into how preprints influence technological innovation and knowledge translation within the open science framework. In this paper, we compare the trend of patent citations and the time taken to be cited in patenting activities across two data sets: preprints versus journal papers, and journal articles with preprint versions versus those without. Additionally, we conduct a comparative analysis to explore the citing behaviors of preprint in patenting process versus journal-based scholarly communications. This analysis addresses the following research questions:

RQ1: How quickly is scientific knowledge from preprints transferred into technological innovation?

RQ2: Do preprints enhance the visibility of early research findings in the technological innovation process?

RQ3: How does the role of preprints differ in terms of accelerating the dissemination and application of scientific knowledge in technological innovation and scholarly communication?

## 2. Related works

*2.1 Knowledge linkage between science and technology: Insights from patent citation analysis*

Since Narin et al. (1997) employed bibliometrics to analyze patent citations to scientific literature, aiming to assess the value of basic research for technological innovation, bibliometric studies of patent citations have become a significant focus in fields such as econometrics, scientometrics, and innovation studies (A.F. van Raan, 2017). The scientific foundation of a patent is acknowledged through curated citations to other patents and research publications to indicate the novelty of a technological invention to the patent office (Alcácer et al., 2009; Ke, 2020; Manjunath et al., 2021). Tracking scientific literature cited in patents reveals a potential pathway through which technological inventors become aware of and recognize scientific articles. This tracking offers valuable evidence regarding the commercial value or, at the very least, the technological innovation impact of scientific research (Jahn et al., 2022; Kousha & Thelwall, 2017; Sternitzke, 2009).

Promoting the integration of interdisciplinary knowledge and the fusion of science and technology is an essential trend for fostering disruptive innovation (Park, Leahey & Funk, 2023; Betz et al., 2023). Patent-to-research citations have been utilized to explore the connection between science and technology among different types of innovation entities and their various innovation activities (Huang et al., 2014). Early studies have demonstrated that scientific advancements are not just theoretical but are actively reshaping industrial technology (Carpenter et al., 1980; Narin et al., 1997; McMillan, Narin & Deeds, 2000). The value of scientific research is increasingly evident in its role in advancing the technological frontier. By exploring the forward citations of two patent groups—those that cite scientific publications and those that do not—researchers have found that knowledge transfer from science correlates with the impact of patents. This suggests that science-based patents tend to have a greater social impact (Wang & Li, 2018; Chen et al., 2024). There is also a growing positive relationship between firms' technological innovation performance and their engagement in basic scientific research activities (Leten et al., 2014; Hung, 2012). Furthermore, studies have shown that the more heavily firms rely on scientific outputs, the more valuable and innovative their producing patents tend to be (Krieger, Monika, & Martin, 2024; Chen et al., 2024).

*2.2 Paper characteristics influence the knowledge transfer from science to technology*

The characteristics of scientific publications also significantly influence their direct technological impact (Ding et al., 2017). Yang (2024) conducted an in-depth analysis of research funded by the National Institutes of Health (NIH) and the National Science Foundation (NSF), demonstrating that the funded research tends to receive higher citation counts in both academic literature and patents. These findings underscore the consistently positive influence of NIH and NSF funding on research impact and dual innovation. Empirical studies further indicate that novel scientific research exhibits greater direct and indirect technological impact, has a broader reach across technology fields, and is more likely to lead to novel patents compared to non-novel scientific research (Veugelers & Wang, 2019). Recently research utilizing large-scale scientific and patent data reveals that basic science papers and novel papers typically take less time to transition into technological applications and receive substantially more patent citations (Ke, 2020; Munari et al., 2022). This finding highlights the importance of considering the nature of scientific output when assessing its potential for technological advancement and innovation. However, the connection between scientific novelty, delayed recognition, and technological impact remains complex (Veugelers & Wang, 2019). A.F. van Raan & J.J Winnink (2018) examined the characteristics of Sleeping Beauties (SBs) cited in patents (SB-SNPRs), suggesting that the average time lag between an SB-SNPR's publication and its first citation in patents is decreasing, and that they are awakened increasingly earlier by a 'technological prince' rather than by a 'scientific prince'.

Patent citation patterns exhibit considerable variation across different technological fields. Analyzing data from patents associated with NIH grants over a 27-year period, Li et al. (2017) found that 30% of NIH-funded research articles are subsequently cited by patents. Notably, disease targeted grants tend to produce publications being slightly more likely to be cited in patents compared to grants that are not disease-oriented, though the difference is marginal. In contrast to physical sciences and engineering, the life sciences exhibit a closer distance between scientific research and technological development (Ahmadpoor & Jones, 2017; Veugelers & Wang, 2019). Regarding the speed of knowledge transfer, Ahmadpoor and Jones (2017) examined 4.8 million U.S. patents and 32 million research articles, revealing an average delay of 6.6 years between the publication of a paper and its citation in a patent application.

*2.3 The impact of Open Access on innovative processes*

While the impact of Open Access (OA) on citation impact and altermatic performance has been extensively studied (Piwowar et al., 2018; Fraser et al, 2020; Wang et al., 2020), its influence on innovation and patent activity remains a relatively underexplored area. Jahn et al. (2022) analyzed a dataset of 215,962 scientific non-patent references (SNPRs) published between 2008 and 2020, identifying a consistent increase in the proportion of OA articles cited in patents, with nearly half of the referenced articles being openly accessible. Their findings also revealed significant variations across countries and disciplines. Notably, patents originating from the US and the UK, as well as those classified under the CPC section "A - Human Necessities," demonstrated a higher uptake of OA literature. Bryan and Ozcan (2021) examined how barriers to the dissemination of academic research affect innovation by analyzing a sample of 43 medical and biotechnology journals published between 2005 and 2012. Their findings indicate that the NIH Open Access (OA) mandate had a modest but measurable positive effect on the citation of SNPRs. Specifically, NIH-funded research saw a 12% to 27% increase in citations following the policy's implementation in 2008. They suggest that the main takeaway for firms is that limited access to research can significantly impact the quality of their innovation. In contrast, making published academic articles freely available online helps reduce barriers to the dissemination of research findings, enhances their broader social impact beyond academia, and particularly facilitates knowledge transfer to industry, thereby fostering a more effective integration of scientific discoveries into technological advancements, ultimately improving innovation efficiency.

Collectively, these studies highlight the potential of OA to bridge the gap between scientific discovery and practical application, offering valuable insights into how open access to scientific knowledge can influence the pace and direction of innovation. However, it remains unclear whether preprint communication can effectively foster the connection between science and technology. Preprints, characterized by early access, open access, and their unreviewed nature, promise to accelerate dissemination in scientific knowledge and increase scientific collaboration. Yet, their potential to facilitate the translation of scientific knowledge into technological innovations and products remains under evaluation. To address this gap, we conduct a comprehensive analysis of the impact of preprints on technology innovation, with a particular focus on the speed at which science is transformed into technology, facilitated by open science communication through the sharing of early research outputs as preprints.

## 3. Data and Methods

Life science, one of the fields that has embraced preprint publishing, with bioRxiv being the most widely used preprint sever, offers a unique environment to explore the aforementioned questions. Several factors contribute to this. First, Life Science is a major research field experiencing significant shifts in research paradigms within the context of open science. The selection of bioRxiv as one of the top ten scientific breakthroughs of 2017 by *Science* (Kaiser, 2017) highlights a substantial cultural shift in academic communication. The COVID-19 pandemic has further accelerated this paradigm shift in the field (Fleerackers et al., 2024). Additionally, previous studies have demonstrated a close relationship between patents and scientific advancements in Life science, as technological development in this field are more directly dependent on the scientific progress (WIPO, 2022).

*3.1 Preprint, articles and patent metadata*

This study uses three data sources: the bioRxiv preprint server for Biology, the Dimension database, and the Altmetric database. The process for identifying patents that cite bioRxiv preprints and for obtaining detailed patent information is illustrated in Fig. 1. We harvested the basic metadata of all preprints submitted to bioRxiv between November 2013 and December 2021 from the Dimension database (N = 144,323). Next, we queried Altmetric.com for mentions of bioRxiv preprints cited by patents with publication date between January 1, 2014 and December 30, 2022. This search yielded a total of 4,060 mentions of 1,803 preprints across 2,934 patents. We didn't identify and combine patent families, as our focus was on pointing the specific patents citing individual preprints rather than the broader patent families. Notably, only 201 out of 2,934 patents (7.74%) were associated with 538 patent families, representing a small percentage.

Additionally, we acknowledge that some preprints may later be published as journal articles, and subsequent patent applications may cite either or both versions. Therefore, we retrieved patent citations for both preprints and their corresponding published versions by querying Altmetric.com via the bioRxiv ID and the journal article DOI, respectively. Besides, we did not distinguish applicant citations and examiner citations. Although these two types of citations may differ in their underlying motivations (Alcacer & Gittelman, 2006; Alcácer et al., 2009) —for instance, for inventors and applicants, citations are made to disclose knowledge of prior art, while for examiners, citations serve as a justification for limiting the claims in the patent application during the review process—prior research has demonstrated a strong correlation between the them (Higham et al., 2021). Therefore, regardless of the type of patent citation, it captures the process of knowledge flow. By combining both examiner and applicant citations, previous studies that have successfully mapped knowledge flow using patent citations (Li et al., 2020; Park et al., 2023). Since the primary objective of this paper is to depict the flow of scientific knowledge through patent citations, merging these two types of citations allows for more comprehensive view of the process.

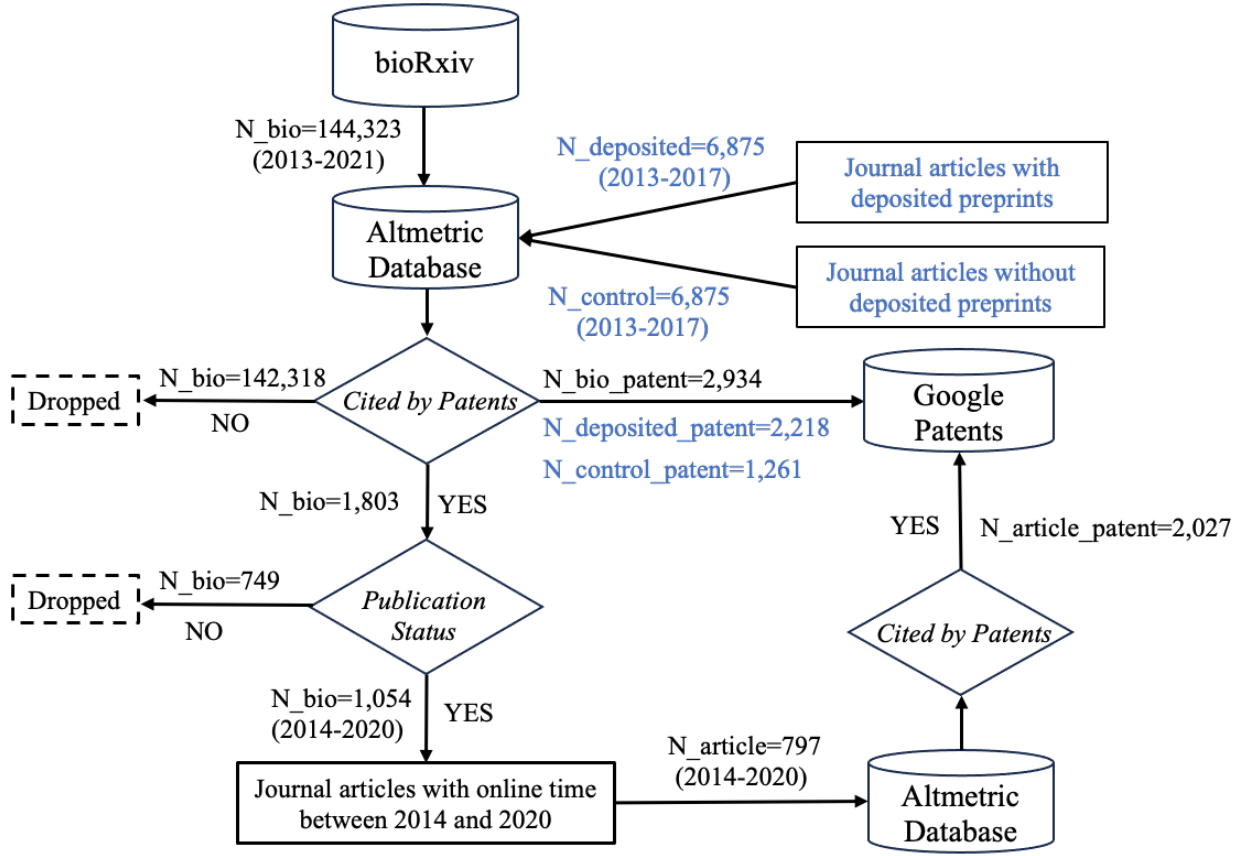

**Fig. 1.** Process for Identifying Patents Citing bioRxiv Preprints and Obtaining Detailed Patent Information. Note: The 6,875 bioRxiv-deposited articles and an equal number of non-deposited control articles were derived from an open analysis dataset originally constructed and validated by Fraser et al. (2020).

*3.2 Identifying COVID-19-related research*

To identify COVID-19-related preprints and articles, we utilized a dataset of preprints labeled as "COVID-19 SARS-CoV-2 preprints from medRxiv and bioRxiv" (https://connect.biorxiv.org/relate/content/181). However, since no pre-existing dataset of patents explicitly marked as COVID-19-related is available, we conducted a keyword-based search in the titles and abstracts of the patents using Python. The keywords included: coro-navirus, coronavirus, covid-19, sars-cov, sars-cov-2, ncov-2019, 2019-ncov, hcov-19, sars-2, pandemic, covid, "Severe Acute Respiratory Syndrome Coronavirus 2", and "2019 ncov". These keywords were optimized by Fleerackers et al. (2024) and have been successfully employed in identifying news mentions of COVID-19 research. To ensure the identified patents were indeed relevant to COVID-19 research, we manually reviewed the results after performing the keyword-based matching, thereby enhancing the accuracy of our analysis.

*3.3 Paired datasets for comparative analysis*

In consideration that the volume of patent application may vary over time in ways that are unrelated to actual changes in patent citations, we established two pairs of experimental and control groups. The first pair consists of bioRxiv preprints and their corresponding journal article versions, while the second includes journal articles with deposited preprints (bioRxiv-deposited article) and those without (control article), with each bioRxiv-deposited article matched to a single random non-deposited article from the same journal, same calendar month and category in the control group. The second dataset is derived from an open analysis dataset originally constructed and validated by Fraser et al. (2020) and is used in a regression analysis to examine the impact of a paper's preprint-deposited status on the difference in patent citations between bioRxiv-deposited articles and control articles. Together, these two datasets provide a robust foundation for our comparative analysis of the effect of preprints on accelerating technological innovation.

*3.4 Regression analysis*

We acknowledge that citation behavior is influenced by a wide range of factors, including document features, author features, journal features and document values (Tahamtan & Bornmann, 2018). To investigate the influence of 'bioRxiv-deposited' status on journal articles' patent citations when controlling for other additional factors, we incorporate a comprehensive set of control variables that significantly influence the citation of articles and are empirically supported in the regression model. The primary independent variable in our model is the 'bioRxiv-deposited' status of the journal paper, coded as a binary variable: "1" for bioRxiv-deposited articles, and "0" for control articles. In addition to this key variable, we include additional control variables related to author, document, and journal characteristics: Country and institutional prestige of the first and last authors, their gender and academic age, number of references, number of pages, online publication time, article open access (*OA*) status, the journal sub-disciplines and impact factor (*IF*). The detailed procedures for data collection and processing of the other variables can be referenced in Fraser et al. (2020), where the journal impact factor (IF) and academic age were calculated using formula (a) and formula (b), respectively:

$$IF_{year} = \frac{Citations_{year-1} + Citations_{year+1}}{Items_{year-1} + Items_{year+1}} \quad (a)$$

$$Academic\ Age = PY_{article} - PY_{first\ pulication} \quad (b)$$

Where *Citations* are sourced from Scopus and *Items* assigned the document type "Article" or "Review" (in Scopus) is included in the denominator. *Items* assigned any other document type are excluded. $PY_{article}$ is the publication year of the article in question, and $PY_{first\ publication}$ is the publication year of the author's first formal publication recorded in Scopus.

It is important to note that, the open access (OA) status of an article is provided by Unpaywall (https://unpaywall.org) and is determined based on its availability, irrespective of the specific format in which it is made openly accessible. While this approach has certain limitations (e.g., some journal articles without associated preprints may have earlier conference papers), the omission of such cases is unlikely to affect the overall findings. A query of Scopus using the DOIs of both bioRxiv-deposited and control articles revealed that none of the articles were classified as 'Conference Papers'. A subsequent query of Web of Science (WoS) yielded 26 bioRxiv-deposited articles and 45 control articles classified as 'Proceedings Paper', representing 0.38% and 0.65% of the respective datasets. Given the minimal proportions, this omission is not expected to impact the results in any meaningful way.

Beyond the aforementioned variables, we incorporate the h-index of the first and last authors, as well as the presence of NIH/NSF funding, as control variables, both of which have been empirically shown to influence research impact (Álvarez-Bornstein & Bordons, 2021; Corsini & Pezzoni, 2023; Yang, 2024). These data are obtained from the SciSciNet, a large-scale open data lake for the science of science research (Lin et al., 2023). The descriptive statistics of the continuous and categorical explanatory variables in the dataset are presented in Table 1 and Table 2, respectively. We collected all patent citation data for these samples. Detailed information on citing patents was obtained from Google Patents, which offers a large searchable index of patents, and providing the full text of patent records, including parent applications, priority applications, legal events, lists of citing patents, family-to-family patent citations, and non-patent literature citations. The bioRxiv API was used to obtain detailed metadata on bioRxiv preprints, including first submission dates and journal publication information. Additionally, the Crossref API was utilized to retrieve the online publication dates of journal articles.

Due to the highly skewed distribution of patent citations, with a variance significantly greater than the mean (8.611 vs. 0.31), and the persistence of this overdispersion even after logarithmic transformation (0.138 vs. 0.086), we employed a negative binomial regression to analyze the patent count data. This method is recommended for analyzing citation count data (Thelwall & Wilson, 2014). We calculated the variance inflation factor (VIF) for all variables in the regression model, and found that all VIF values were below the threshold of 10, indicating that multicollinearity is within an acceptable range. We initially performed a reduced regression model to assess the influence of the 'bioRxiv-deposited' status in the absence of the other independent variables listed in Table 1 and Table 2. Following this, we applied a full regression model, including all variables.

**Table 1** The summary statistics of the continuous explanatory variables

| Variables | bioRxiv-deposited | | | | Control | | | |
| --- | --- | --- | --- | --- | --- | --- | --- | --- |
|  | Min | Max | Mean | Median | Min | Max | Mean | Median |
| Author count | 1 | 402 | 6.96 | 5 | 1 | 484 | 7.91 | 6 |
| First author academic age | 0 | 22 | 5.90 | 5 | 0 | 22 | 6.10 | 5 |

| | | | | | | | |
|---|---|---|---|---|---|---|---|
| Last author academic age | 0 | 22 | 15.52 | 17 | 0 | 24 | 16.12 | 19 |
| First author h-index | 0 | 120 | 12.68 | 10 | 0 | 181 | 11.79 | 8 |
| Last author h-index | 0 | 257 | 40.24 | 34 | 0 | 225 | 37.58 | 31 |
| Journal impact factor | 0.21 | 37.19 | 6.29 | 4.53 | 0.21 | 37.19 | 6.29 | 4.53 |
| Number of pages | 2 | 202 | 13.84 | 12 | 2 | 116 | 13.16 | 12 |
| Number of references | 2 | 488 | 55.82 | 53 | 1 | 443 | 54.33 | 51 |
| Online publication year | 2013 | 2017 | 2016.41 | 2017 | 2013 | 2017 | 2016.41 | 2017 |

**Table 2** The summary statistics of the categorical explanatory variables

| Groups | First author is female | Last author is female | First author is from USA | Last author is from USA | First author is from top100 institute | Last author is from top100 institute | Article is OA | NIH | NSF | Life Sciences | Health Sciences | Physical Sciences |
|---|---|---|---|---|---|---|---|---|---|---|---|---|
| bioRxiv-deposited | 1,979 | 1,191 | 3,391 | 3,403 | 1,843 | 1,809 | 6,129 | 2445 | 942 | 6,193 | 1788 | 1435 |
| Control | 2,410 | 1,552 | 2,510 | 2,519 | 1,156 | 1,146 | 5,839 | 1863 | 501 | 6,193 | 1788 | 1435 |

Note:'NIH': Supported by grants from NIH; 'NSF': Supported by grants from NSF.

## 4. Results

*4.1. Increasing citations to bioRxiv preprints in patents*

Fig. 2a reveals a clear trend of a steady increase in the number of biorRxiv preprints cited in patents over time, although the corresponding percentage (indicated within brackets in Fig. 2a) remains relatively low. The observed decline in the number of patent-cited preprints in 2021 is likely due to the lag between the publication of an article and the application for a patent, with an average 18-month between filing a patent application and its public availability through publication (WIPO, 2022). Consequently, the citation window ending in 2022 might not yet provide a complete picture for patent-cited bioRxiv preprints posted in 2021 due to the time required for patent information to become publicly available.

There is a sharp rise in the number of patent-cited bioRxiv preprints in 2020, followed by a rapid decline in 2021 (cf. Fig. 2a). Further analysis reveals that 333 and 103 patent-cited bioRxiv preprints in 2020 and 2021, respectively, are related to COVID-19, accounting for nearly half of the total patent-cited bioRxiv preprints during these two years. This trend is understandable given significant research and innovation efforts to combat the SARS-CoV-2 virus and the COVID-19 disease, which began in late 2019. The rapid communication model enabled by preprints has also contributed to accelerating research and innovation in the field of biology. The 406 COVID-19-related preprints are cited 920 times by 562 patents, of which 515 patents are related to COVID-19 detection technology, vaccine technology or treatment technology. The explosive increase in the number of bioRxiv preprints in patents suggests that COVID-19-related knowledge is being transferred from preliminary research into innovations in coronavirus technologies at unprecedented pace during the global crisis.

To differentiate the influence of COVID-19, we divided the preprints into two groups based on their submission dates: the pre-COVID-19 era (2014–2019) and the COVID-19 era (2020). This division aligns with the WHO's first use of the term "2019-nCoV" to describe the novel coronavirus on January 10, 2020 (WHO, 2020). For both groups, we collected their citing patents with publication dates up until the end of 2022. In the first group, we identified 2,122 patent citations from 1,097 patents referencing 839 bioRxiv preprints. In the second group, we identified 1,588 patents citations from 1,645 patents referencing 715 preprints. Fig. 2b shows the distribution of patent citations for two groups of bioRxiv preprints, revealing a distinctly skewed pattern. Overall, among preprints with non-zero patent citations, more than 60% received only a single citation, while approximately 20% of preprints received at least three patent citations. This skewed trend was especially pronounced before 2019, with over 67% of patent-cited preprints receiving just one citation, and fewer than 16% receiving more than three patent citations.

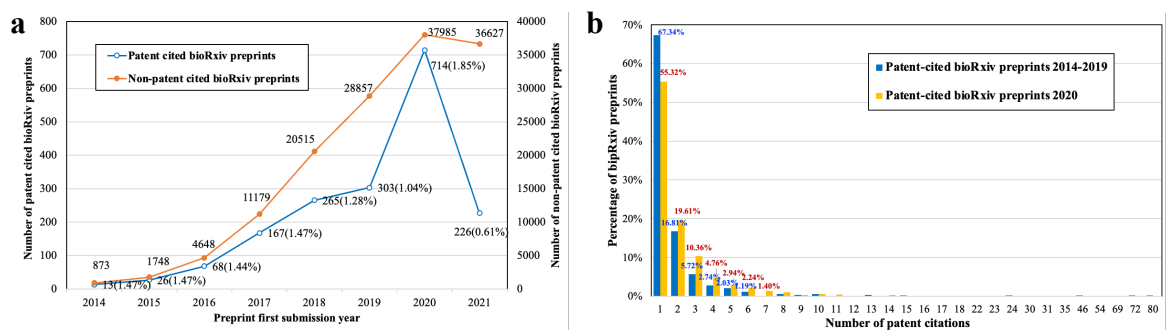

**Fig. 2.** Patent citations to bioRxiv preprints. **(a)** Number of patent-cited preprints vs. none-patent cited papers over year. **(b)** Distribution of patent citations to bioRxiv preprints 2014–2019 vs. bioRxiv preprints 2020.

Furthermore, we found that the ratio of patent-cited bioRxiv papers to the total number of bioRxiv papers is much higher than that of patent-cited arXiv papers with a first submission year between 2014 and 2021 (1.24% vs. 0.68%)[1]. This indicates that technological innovation in the field of biology is more heavily dependent on academic research than in the fields of physics, mathematics and computer science, which are the predominant fields covered by arXiv. However, it is important to note that bioRxiv preprints across different categories exhibit varying levels of patent citation performance (Fig. 3).

---

[1] We queried Altmetric.com for mentions of arXiv preprints cited by patents between January 1, 2014 and December 30, 2022 at the same, 16,326 mentions of 13,548 preprints across 11,050 patents for arXiv preprints.

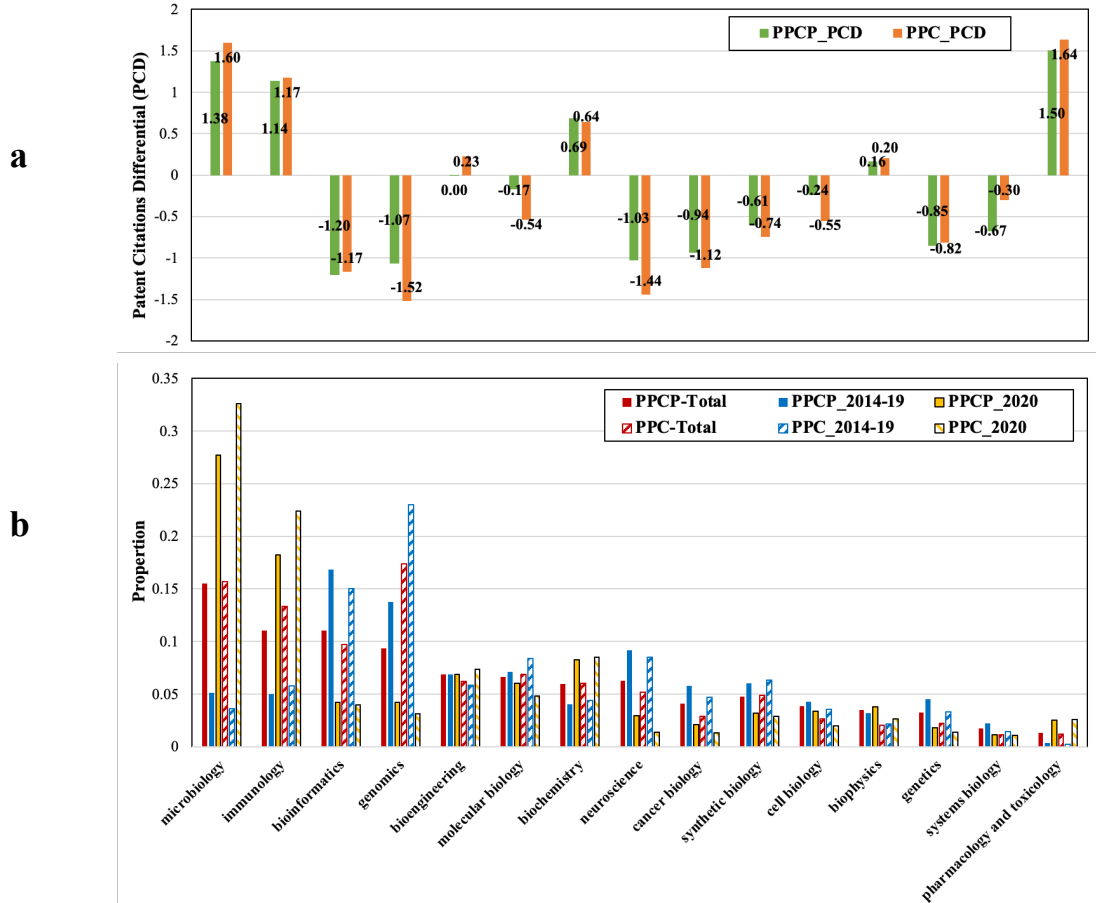

$$Proportion\ of\ patent\ cited\ preprints (PPCP) = \frac{Number\ of\ patent\ cited\ preprints\ in\ the\ category}{Total\ number\ of\ the\ patent\ cited\ preprints}$$

$$Proportion\ of\ patent\ citations (PPC) = \frac{Number\ of\ patent\ citaitons\ for\ preprints\ in\ the\ category}{Total\ number\ of\ patent\ citaitons\ for\ all\ preprints}$$

$$Differential\ of\ PPCP (PPCP\_PCD) = 2 \times \frac{PPCP_{2020} - PPCP_{2014\_2019}}{PPCP_{2020} + PPCP_{2014\_2019}}$$

$$Differential\ of\ PPC (PPC\_PCD) = 2 \times \frac{PPC_{2020} - PPC_{2014\_2019}}{PPC_{2020} + PPC_{2014\_2019}}$$

**Fig. 3.** The TOP15 categories with the highest amount of patent-cited bioRxiv papers. (**a**) Proportion of number of patent-cited bioRxiv preprints (*PPCP*) and their patent citations (*PPC*) within each category. (**b**) Patent citations differential (*PCD*).

For bioRxiv papers cited in patents during 2014 and 2022, the top four most frequently cited categories are microbiology (15.5%), immunology (11.1%), bioinformatics (11.1%) and genomics (9.4%) (see Fig. 3). When considering the proportion of patent citations for papers within each category (*PPC*), the fields of genomics (17.4%), microbiology (15.7%), and immunology (13.3%) stand out prominently. Furthermore, we examined the proportion of patent-cited bioRxiv preprints (*PPCP*) and their associated patent citations (*PPC*) across two time periods: 2014–2019 and 2020. The differentials in *PPCP* and *PPC* between the two groups were calculated using the formulas *PPCP_PD* and *PPC_PD*, respectively (see Fig. 3). This comparative analysis reveals that the microbiology field exhibits the most significant shifts in patent citations before and after 2020, with *PPCP_PD* and *PPC_PD* values of 1.38 and 1.6, respectively, along with the highest *PPCP* and *PPC* values among all fields. A similar trend is observed in the immunology field. Additionally, although the pharmacology and toxicology field has relatively lower *PPCP* and *PPC* values, it demonstrates a notable increase in both patent-cited preprints and patent citations after 2019, with *PPCP_PCD* and *PPC_PCD* values of 1.5 and 1.64, respectively. Notably, the bioengineering filed maintains a strong connection with patents across both periods, highlighting its consistent relevance to technological innovation.

*4.2. Accelerating the knowledge transfer from science to technology*

The speed at which scientific knowledge is transferred into the patenting process is a crucial factor in understanding the time lag between scientific discovery and its corresponding invention. This lag is typically defined as the time interval between the publication of a paper and the year it is cited in a patent application. How long does it take for patents to recognize preprints, and does this time lag differ when compared to journal articles? To explore this question, we examined the time lag between patents and preprints, comparing it to the time lag for their journal versions.

Our analysis reveals a significant acceleration in the time lag between the submission of a preprint and its subsequent citation in a patent, particularly during the COVID-19 pandemic. In this study, the time lag is measured as the duration between the preprint's first submission date and the patent's filing (application) data. Additionally, we analyzed the time lapse between the preprint's submission date and the first patent citation date. Figure 4a illustrates the evolving pattern of patent citations across two distinct periods, highlighting the shifting dynamics of knowledge transfer. A striking divergence emerges when comparing preprints submitted between 2014 and 2019 with those submitted in 2020. The data indicates that preprints posted during the COVID-19 pandemic were recognized by patents at an accelerated pace, with an average time lag of just 10.78 months and a median of 17 months. Remarkably, over 80% of patent citations occurred within 15 months of preprint submission. This is in stark contrast to preprints from the 2014–2019 period, which experienced a significantly longer knowledge transfer timeline. The average and median time lags were 16.62 months and 20 months, respectively, with fewer than 45% of citations occurring within 15 months. Moreover, it took approximately 26 months for preprints to accumulate 80% of patent citations, reflecting the relatively slower integration of preprints into innovation pipeline in the pre-pandemic era.

We further examined the time taken for each preprint to receive its first patent citation (see Fig. 4b). Notably, preprints can be cited by patents almost immediately after submission. For preprints submitted prior to 2020, approximately 2.17% were cited by patents within the same month they were posted, and over 80% received their first patent citation within 24 months of submission. However, during the COVID-19 pandemic, this timeline contracted significantly. The proportion of bioRxiv preprints cited within the same month of posting rose to 6.9%, demonstrating an increased urgency in integrating cutting-edge research into patent applications. The acceleration is further evident in the fact that most preprints received their first patent citation approximately 12 months after submission, with over 80% achieving this milestone within just 14 months.

This rapid reduction in time lag highlights a notable shift in how scientific knowledge is transferred into patentable innovations during the pandemic era. Preprint communication lowers traditional barriers in the knowledge transfer process, while the urgency of addressing the pandemic likely motivated patent innovators to accelerate the integration of emerging research—rapidly disseminated and freely accessible through preprints—into their inventions. This trend has facilitated a more immediate and efficient translation of scientific research into technological and commercial applications.

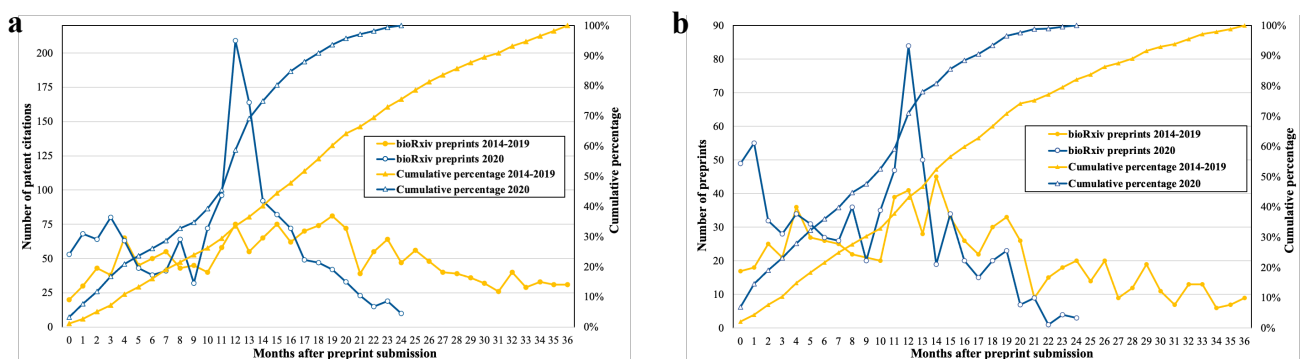

**Fig. 4.** Trends in patent citations to bioRxiv preprints submitted between 2014–2019 and 2020. **(a)** Time lag between the filing date of patents and the submission date of preprints. **(b)** Time lag between the shortest filing date of patents and the submission date of preprints.

BioRxiv preprints tend to be recognized by patents more quickly than their corresponding journal articles. To explore this further, we analyzed patent citations to both preprints and their corresponding journal versions. Of the 1,553 bioRxiv preprints cited by patents between 2014 and 2020, 1,054 were eventually published in journals (referred to as 'article' hereafter), with 797 of these articles having online publication dates within the 2014-2020 timeframe. To gather patent citations data, we queried Altmetric.com for patent citations to these 797 articles, covering the period from January 1, 2014, to December 30, 2022. This search identified 3,026 mentions of 388 articles across 2,027 patents, meaning about 48.68% of the 797 articles were cited by patents. In comparison, there were 1,545 mentions of the corresponding 388 bioRxiv preprints across 1,071 patents. The average number of patent citations for preprints and journal versions is 5.98 and 7.79 respectively, with median value of 2 and 3 respectively. Additionally, a Spearman regression analysis revealed a weak but statistically significant correlation of 0.33 (*P*<.001), between patent citations to preprint and their journal articles. These findings suggest that while peer-reviewed articles are more frequently recognized and pursued by patent applicants than their preprint versions, it's not always the case that an article will be commercially developed if its preprint has been.

The aging patterns of patent citations for bioRxiv preprints and journal articles exhibit distinct trends (cf. Fig. 5a). BioRxiv preprints experience a rapid rise in patent citations shortly after submission, peaking around the fourth month, followed by a decline. However, they undergo a second surge between the 10th and 13th months. In contrast, patent citations to journal versions exhibit a relatively slower aging trend, with the peak occurring around the 13th month. The aging rates of the first patent citations (see Fig. 5b) also reveal significant differences. Preprint versions display a bimodal trend, while journal versions show a slower aging process, with patent citations declining starting 13 months after publication.

Furthermore, when considering the time required for preprints to undergo peer review and be formally published in journals, our dataset reveals an average time lag of approximately 8.18 months between the submission of preprints and the online publication of their corresponding journal versions. These findings suggest that preprints play a crucial role in accelerating the transfer of new knowledge to technological innovation.

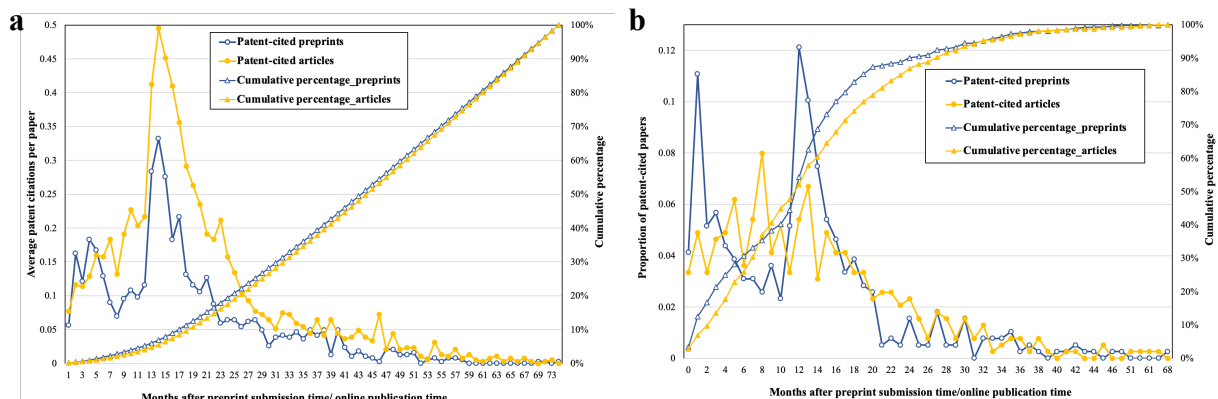

**Fig. 5.** Trends in patent citations to bioRxiv preprints and their published journal articles. **(a)** Time lag between the filing date of patents and the online publication time of articles/submission time of preprints. **(b)** Time lag between the filing date of first citing patents and the online publication time of articles/submission time of preprints.

To gain a deeper understanding of the evolving role of preprints in accelerating knowledge transfer from science to technology and their impact on the scientific communication landscape, we conducted a comparative analysis of the patent assignees citing the two types of papers: bioRxiv preprints and their corresponding journal articles. Fig. 6 shows the proportion of patent assignees citing preprints and journal articles across various types of institutions, including companies, research institutes, universities, hospitals, government organizations to explore how different institutions engage with the two types of scientific knowledge.

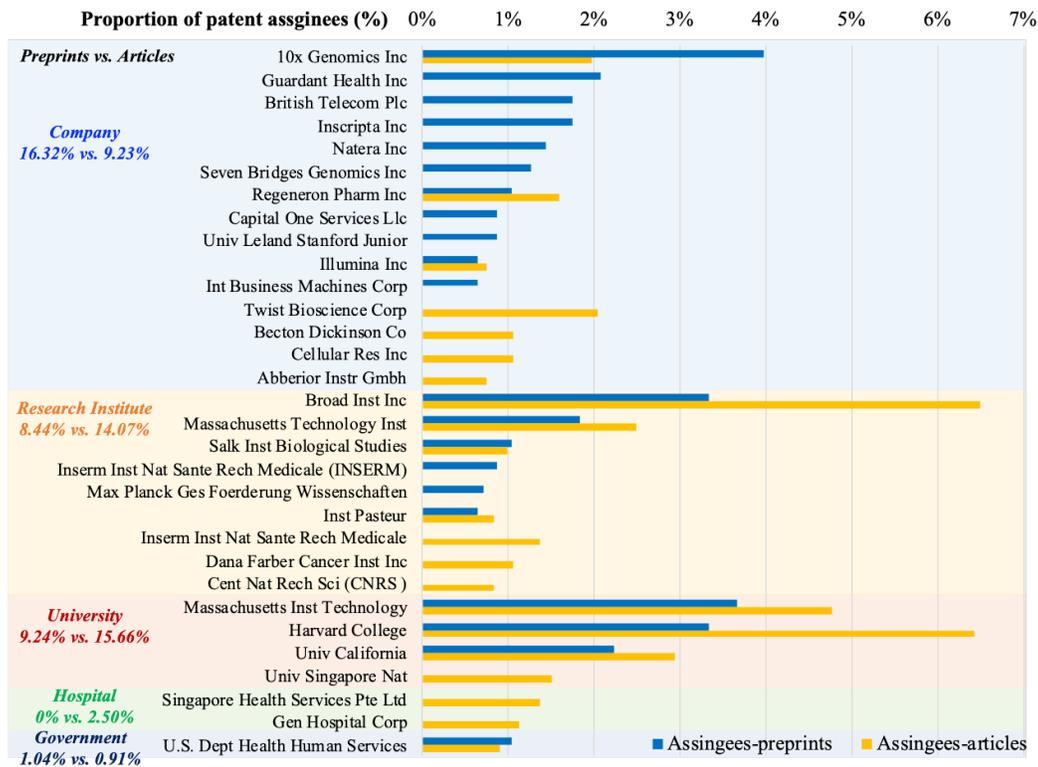

**Fig. 6.** Comparison of top 20 patent assignees citing preprints and journal articles. The two bars for each institution represent citations of preprints (blue) and journal articles (yellow).

Overall, biotechnology companies such as 10x Genomics, Guardant Health, Seven Bridges Genomics and Inscripta cite preprints more frequently than journal articles, suggesting that companies in the biotechnology and pharmaceutical sectors rely more heavily on preprints to early access latest research findings, and track cutting-edge research, which helps accelerate the pace of innovation. While research institutes (e.g., Broad Institute) and universities (e.g., MIT, Harvard College, and University of California) also draw substantial knowledge from preprints, they show a strong preference for citing peer-reviewed journal articles, especially if the preprints already have corresponding journal versions, highlighting the importance of validated and reliable peer-reviewed science in academic and research institutions. In contrast, hospitals such as General Hospital Corporation and Singapore Health Services focus exclusively on citing journal articles. This reliance on peer-reviewed publications highlights the critical need for rigorous and validated findings in clinical and healthcare settings, where accuracy and reliability are paramount. Government agencies, such as the U.S. Department of Health and Human Services, cite both preprints and journal articles in their patent filings, though their overall proportion of citations remains low. This is understandable, as government bodies are not central players in the technological innovation chain, which is often driven by private companies and research institutes.

*4.3. Enhancing the visibility of early research results in technological innovation process*

Preprints serve as an important channel for authors to rapidly disseminate their work and enhance its rigor, with most preprint authors eventually publishing in peer-reviewed journals (Fraser et al., 2020). But when, how and to what extent does sharing early research results, such as preprints, influence the transfer of science knowledge to technological innovation? To address these questions, we first investigated the aging patterns of patent citations to 797 published bioRxiv preprints and their corresponding journal articles. We then conducted a comparative analysis of patents citations between journal articles with and without associated bioRxiv preprints.

The patent citation trends for both bioRxiv preprints and their corresponding journal articles (yellow curve in Fig.7), and for bioRxiv preprints alone (blue curve in Fig. 7), exhibit similar aging patterns, with a significant decline occurring approximately 12 months after the online publication. However, patent citations to preprints demonstrate a faster aging rate (blue curve with hollow triangular markers in Fig.7), particularly within 10~12 months after the online publication. This faster aging may be attributed to the early visibility and accessibility of preprints. As the knowledge from preprints is subsequently published in journals, it continues to be transferred from science to technology (the yellow with solid circular markers in Fig. 7), highlighting that preprints enable the early sharing of knowledge and facilitate its transfer from science to technology.

Notably, most preprints continue to be cited even after their publication in journals (blue curve in Fig.7). Nearly 70% of patent citations occur after the preprint has been published in a journal, revealing that the citation trends of preprints in patents differ significantly from the citation behavior of journal article authors toward preprints. The latter behavior exhibits a significant decline in average citation rates following the online publication of the journal article (cf. Figure 6b in Wang, Chen, & Glänzel, 2020; Figure 5 in Fraser et al., 2020). A possible explanation for this phenomenon might be that the limited access to journal databases, especially among firms, including small- and medium-sized enterprises (SMEs) without collaborators at universities or public research institutions (Jahn et al., 2022; Leten et al., 2014). Unlike academic researchers, who often have institutional access to subscription journals, many enterprises—especially SMEs—do not purchase journal databases and thus cannot access academic outputs in subscription-based journals. This difference in citation behavior underscores the unique role of preprints in bridging the gap between academic research and industrial innovation. While academic authors prefer to cite the journal version rather than the preprint version when both are available (Wang, Chen, et al., 2020), industry inventors appear to value the "open access" and "early access" of preprints. This finding emphasizes the significant role of preprints as a valuable resource for innovation, particularly in industries where access to traditional subscription-based academic resource is limited.

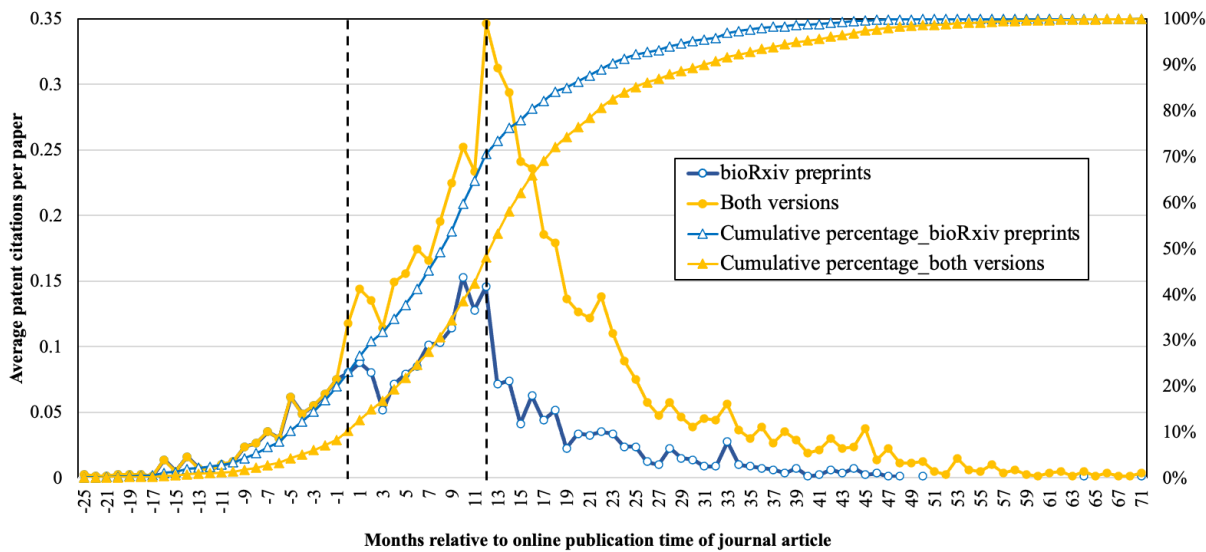

**Fig. 7.** Aging pattern of patent citations based on online publication time. The yellow curve represents patent citations to both bioRxiv preprints and their corresponding journal articles, and the blue curve represents patent citations exclusively to bioRxiv preprints.

The aging pattern of patent citations of both versions (the yellow curve in Fig. 7) demonstrates that sharing early research results, such as preprints, accelerates the transfer of knowledge to technological innovation. Moreover, the subsequent publication of peer-reviewed journal articles continues to play a crucial role in facilitating this transformation, often having an even greater impact. This is particularly evident when considering that the average number of patent citations to the journal articles is significantly higher than to the corresponding preprint versions. (cf. Fig. 4a & Fig. 5 yellow curve with solid circular markers). These findings suggest that preprints increase the visibility of early research results in the patent process, thereby strengthening the connection between science and technology.

To further assess the impact of preprints on technological innovation, we conducted a comparative analysis of patent citations between two groups of articles: one comprising journal articles with a corresponding preprint version deposited in bioRxiv, referred to as 'bioRxiv-deposited' articles, and the other consisting of non-deposited articles serving as a control group. For this analysis, we utilized an open dataset originally constructed and validated by Fraser et al. (2020), which includes 6,875 bioRxiv-deposited articles and an equal number of non-deposited control articles. These control articles were carefully selected to be broadly similar to the bioRxiv-deposited group, thereby minimizing the influence of article characteristics on impact performance. This dataset provides a robust foundation for our comparative analysis of the effect of preprints on accelerating technological innovation. Subsequently, we queried Altmetric.com to identify patent mentions of both the bioRxiv-deposited articles and the control articles by patents between January 1, 2014 and December 30, 2022. The results (Table 3) reveal a significant trend:

among the bioRxiv-deposited articles, 612 were cited in 2,218 patents, accumulating a total of 2,887 citations. In contrast, the control group, comprising 396 articles, received 1,391 citations across 1,261 patents. This discrepancy indicates that bioRxiv-deposited articles were cited in patents 1.55 times more frequently than those in the control group. Additionally, the bioRxiv-deposited articles exhibited a lower frequency of zero patent citations compared to the control group and this pattern is consistent across all years.

In the comparative analysis of the mean of patent citations per paper between bioRxiv-deposited articles and control articles, we applied a log transformation to the patent citation counts (*LCPP*) to normalize the data (Thelwall, 2017) and to reduce the influence of extremely highly cited papers on the results. *LCPP* was calculated as the mean of the log-transformed patent citation counts for all articles within each group. Then, in order to quantify and measure the patent citation differential of bioRxiv-deposited articles versus control articles, we used the optimized function, Impact Differential Ratio (IDR) (Wang, Chen, et al., 2020), to calculate the difference in patent citations (*P_IDR*) between the two sample groups. The two formulas as:

$$LCPP = \frac{1}{n}\sum_{i=1}^{n} \log(Patent\ Citation_i + 1)$$

$$P\_IDR = 200 \times \frac{LCPP_{bioRxiv\_deposited} - LCPP_{control}}{LCPP_{bioRxiv\_deposited} + LCPP_{control}}$$

Where *n* represents the total number of articles within each group; $LCPP_{bioRxiv\_deposited}$ and $LCPP_{control}$ denote the mean of the log-transformed patent citation counts for bioRxiv-deposited articles and control articles, respectively.

The bioRxiv-deposited articles generally receiving a higher mean of patent citations per year, as reflected in the *P_IDR* values (the last column of Table 3). The most notable increase occurs in 2017, suggesting that bioRxiv-deposited papers attract a significant number of early patent citations. This observation points to a potential acceleration in the recognition and utilization of research disseminated via preprints in patent filings. We will further investigate this hypothesis in the following analysis using negative binomial regression.

**Table 3** The description statistics of patent citations to bioRiv-deposited articles and control articles

| Year | Number of articles | bioRxiv-deposited | | | Control | | | P_IDR |
| --- | --- | --- | --- | --- | --- | --- | --- | --- |
| | | Number of zero patent citations (Percentage) | Sum of patents | LCPP | Number of zero patent citations (Percentage) | Sum of patents | LCPP | |
| 2013 | 1 | 1 (100%) | 0 | 0 | 1 (100%) | 0 | 0.00 | / |
| 2014 | 261 | 222 (85.06%) | 378 | 0.229 | 231 (88.51%) | 157 | 0.154 | 39.3 |
| 2015 | 772 | 691 (89.51%) | 324 | 0.127 | 703 (91.06%) | 373 | 0.123 | 2.99 |
| 2016 | 1710 | 1571 (91.87%) | 703 | 0.099 | 1622 (94.85%) | 297 | 0.056 | 56.25 |
| 2017 | 4131 | 3778 (91.45%) | 1480 | 0.105 | 3922 (94.94%) | 564 | 0.055 | 61.47 |
| Total | 6875 | 6263 (91.10%) | 2885 | 0.110 | 6479 (94.24%) | 1391 | 0.067 | 49.25 |

To ensure a fair comparison of patent citations across articles published at different times and to mitigate potential bias in the regression analysis, we controlled the citation window to a fixed period of 60 months. This choice is justified by two key considerations: first, the longest available citation window for articles published in 2017 is 60 months; second, more than 80% of patent citations occur within this period, making it a representative and meaningful timeframe for analysis. The results indicate that 585 bioRxiv-deposited articles were cited in 2,019 patents, accumulating a total of 2,589 citations. In contrast, the control group, 383 articles received 1,136 citations across 1,261 patents. The distribution of patent citations for each group is highly skewed (Fig. 8a). More than 46.9% of articles received only one citation from patents, while fewer than 31.5% of bioRxiv-deposited articles and 27.2% of control articles received at least three patent citations.

Notably, the bioRxiv-deposited articles showed a clear advantage in patent citations following their online publication (Fig. 8b), with the value of the paired difference ($P\_IDR$) ranging from 10.44 to 118.88. Specifically, the cumulative $P\_IDR$ value for the 60-month citation window is 48.69, while for shorter citation windows, the values remain consistently high (49.47 for 48 months, 50.01 for 36 months, 47.85 for 24 months, and 51.09 for 12 months). To determine whether the difference in patent citations between bioRxiv-deposited articles and control articles is statistically significant, we conducted a nonparametric two-sided paired Wilcoxon signed-rank test. This test is well-suited for our dataset as it does not require a normal distribution of the differences between paired observations, making it robust for dataset with considerable number of zero values. The Wilcoxon signed-rank test revealed a statistically significant patent citation advantage for bioRxiv-deposited articles over non-deposited articles across all citation windows. Specifically, the test yielded $Z$-values of -6.635 ($p < 0.001$), -6.530 ($p < 0.001$), -5.902 ($p < 0.001$), -4.520 ($p < 0.001$), and -3.317 ($p < 0.001$) for the 60-month, 48-month, 36-month, 24-month, and 12-month citation windows, respectively, confirming that preprint deposition is associated with a patent citation advantage over time.

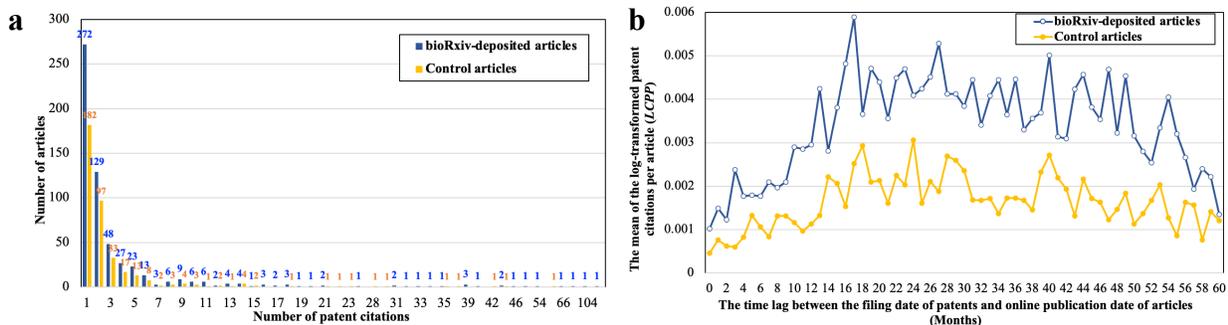

**Fig. 8.** Comparative analysis of patent citations to bioRxiv-deposited articles and control articles. (a) The distribution of patent citations. **(b)** Aging pattern based on online publication time.

The above analysis demonstrates that the patent citation advantage of bioRxiv-deposited articles emerges as early as the first year post-publication and remains consistently strong across all citation windows. This sustained impact suggests that preprint deposition not only enhances initial visibility but also contributes to long-term technological impact, persisting for at least five years. By providing an accessible and early source of scientific information for inventors, the preprint versions of journal articles enable patented technologies to incorporate cutting-edge research more rapidly than traditionally published journal articles alone. However, as noted in our Data and Methods section, summary statistics for factors related to document features, publication venues and authorship (Tables 1 and 2) reveal some key differences between bioRxiv-deposited articles and control articles. To accurately assess the impact of a journal article's bioRxiv-deposited status on its patent citation advantage, it is necessary to control for other potentially influential variables listed in Tables 1 and 2.

We performed negative binomial regression analysis on patent citation counts for the 12-month, 24-month, 36-month, 48-month and 60-month citation window, respectively. The results of are summarized in Table 4, with full details provided in Supplementary Tables A1–A5. The analysis proceeded in two stages: first, a reduced model was estimated, incorporating only the primary predictor variable, 'bioRxiv deposited' status (labeled as 'bioRxiv-deposited' in Table 4)". Subsequently, a full model was fitted, incorporating all covariates detailed in Tables 1 and 2, to account for additional factors potentially influencing the dependent variable. Incidence rate ratios (IRR) were derived by exponentiating the regression coefficients （$\exp(\beta)$）, providing an interpretation of the relative change in the outcome (i.e., patent citation counts) for a one-unit increase in the corresponding predictor. For instance, an IRR of 1.151 for the predictor "IF" on the dependent variable "patent citations" indicates that for every unit increase in IF, an article is expected to receive 1.151 times as many patent citations.

Results (Table 4) from the reduced regression models align with those from the initial analysis, confirming that articles deposited to bioRxiv accrue a significantly higher number of patent citation across all time windows, with an IRR$_{reduced\ model}$ of 2.098 at 12 months and 2.073 at 60 months. Moreover, the full regression models further substantiate this finding, demonstrating that even after controlling for variables associated with document features, publication venue and authorship, the 'bioRxiv-deposited' status remains a significant independent predictor of patent citations, with IRR$_{full\ model}$ between 1.696–1.782, though slightly lower than in the reduced model.

Previous studies have demonstrated that bioRxiv papers have advantages in terms of both citations and Altmetric counts (Fraser et al., 2020). Our results reveal that bioRxiv papers also possess a patent citation advantage, further enhances the positive effect of preprints on technological innovation and facilitating the transforming of scientific theories into practical applications.

**Table 4** The results of the influence of the 'bioRxiv-deposited' status for the patent citations from reduced and full regression model. (see Supplementary Tables A.1–A.5 for full regression results)

| Outcome variables | bioRxiv-deposited | | Control | | IRR$_{reduced\ model}$ (95%CI) | Std. Error | IRR$_{full\ model}$ (95%CI) | Std. Error |
| --- | --- | --- | --- | --- | --- | --- | --- | --- |
| | Mean | Std. Error | Mean | Std. Error | | | | |
| 12-month patent citations | 0.07 | 0.818 | 0.04 | 0.407 | 2.098*** (1.795-2.452) | 0.0796 | 1.740*** (1.451-2.09) | 0.0929 |
| 24-month patent citations | 0.16 | 1.476 | 0.08 | 0.702 | 2.056*** (1.844-2.291) | 0.0553 | 1.696*** (1.492-1.927) | 0.0655 |
| 36-month patent citations | 0.24 | 2.152 | 0.12 | 1.021 | 2.104*** (1.923-2.305) | 0.0462 | 1.696*** (1.52-1.891) | 0.0557 |
| 48-month patent citations | 0.32 | 2.727 | 0.15 | 1.311 | 2.081*** (1.921-2.257) | 0.0411 | 1.754*** (1.59-1.937) | 0.0503 |
| 60-month patent citations | 0.38 | 3.154 | 0.18 | 1.575 | 2.073*** (1.923-2.234) | 0.0384 | 1.782*** (1.624-1.956) | 0.0477 |

Note: *** < .001

*4.4. Dissemination speed of preprints in technological innovation differs from that in scholarly communication*

The scholarly community has embraced preprints for several reasons, including early discovery, open access, and early feedback (Maggio et al., 2018). Additionally, academic researchers use preprints to increase their scholarly impact, such as citations (Wang, Glänzel, et al., 2020), and social impact, such as shares on social media platforms (Fraser et al., 2020; Wang, Chen, et al., 2020), as well as mentions in patents, which can reflect the potential innovation and economic value of their research results.

To better understand the similarities and differences in the usage and dissemination of preprints in technological innovation versus scholarly communication, we conducted a comparative analysis of the aging trends of citations from both articles and patents to bioRxiv preprints (cf. Fig. 9). The results show that the average number of patent citations per preprint is significantly lower than the number of article citations.

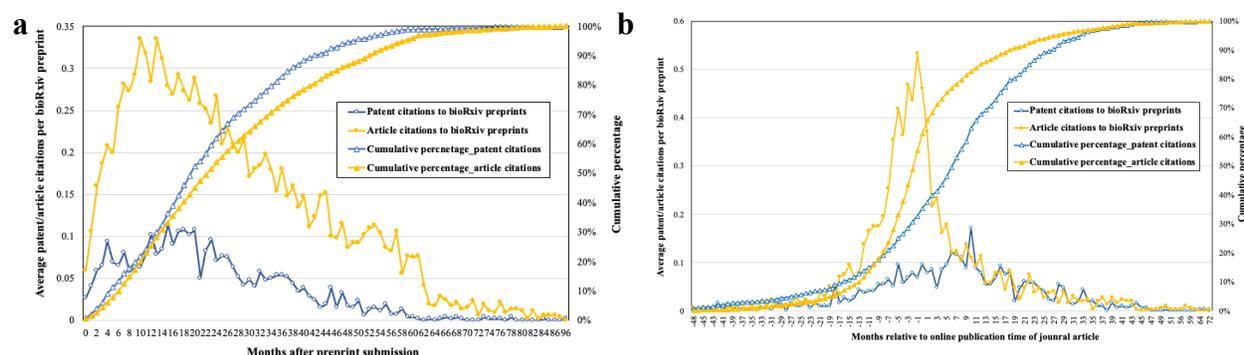

**Fig. 9.** Aging patterns of patent citations and article citations to preprints. **(a)** Comparison between the aging patterns of patents citations and article citations to preprints based on preprint submission time. The dataset includes the 520 patent-cited bioRxiv preprints that also have been received at least one citation from COCI (OpenCitations Index of Crossref open DOI-to-DOI references, Version 11 released on 2021-09-03)**.** **(b)** Comparison between the aging patterns of patents citations and article citations to preprints based on online publication time. The dataset includes the 291 out of the above 520 patent-cited bioRxiv preprints, which have journal article versions.

Interestingly, the aging trends are similar between the two categories. However, there is a notable distinction in the aging and obsolescence patterns of preprints cited in patents compared to those cited in research articles. For instance, approximately half of the citations from patents accumulate within 19-20 months, whereas it takes about 23 months for citations from articles to reach the same level (cf. Fig. 9a). This suggests that the dissemination speed of preprints in technological innovations is faster than in scholarly communication, although it is not as widespread.

We further investigated the aging pattern of patent citations based on their online publication dates, and compared with that of article citations to preprints. Significant differences were observed between the two aging trends. Citations to preprints become more frequent in the months leading up to the journal's online publication and then sharply decline afterward (cf. Fig. 9b yellow line). This trend aligns with previous studies (Fraser et al., 2020; Wang, Chen, et al., 2020), which found that authors prefer to cite the journal version rather than the preprint version when both are available. By contrast, this decline in patent citations occurs approximately 10 months later (cf. Fig. 9b, the blue line).

In addition, 75.6 % of the 291 bioRxiv preprints were cited before their journal publication, accounting for 55.4 % of the total 2,363 article citations. For patent citations, 40.89 % preprints were cited before journal publication, representing 32.7 % of all 1,052 patent citations (cf. Fig. 9b blue line). The substantial number of post-online-publication patent citations indicates an ongoing trend where preprints continue to be cited in patents even after their journal publication. This underscores the importance of recognizing the value of open science and the "open access" effect of preprints, which extend the impact of scientific research on industrial innovation and social economy.

## 5. Conclusion and discussion

To better understand the role of preprints in facilitating the transfer of scientific knowledge to technological innovation within the Open Science ecosystem, we conducted a comprehensive scientometric analysis of patent citations to bioRxiv preprints. This multi-dimensional analysis aims to measure and assess the impact of preprints on accelerating knowledge transfer from science to technology, highlighting the crucial role preprints play in linking scientific research to technological advancements.

There is a consistent and notable upward trend in the number of preprints cited in patents over time, although the overall percentage remains relatively low. Particularly significant is the sharp increase in the number of patent-cited bioRxiv preprints in 2020, followed by a rapid decline in 2021, largely due to the impact of the COVID-19 pandemic. Preprints play a crucial role in accelerating the transfer of new knowledge to technological innovation. The time lag between patents and preprints is relatively short and has decreased further during pandemic.

Preprints play a critical role in accelerating innovation, while journals remain essential for academic rigor and reliability. BioRxiv preprints tend to be recognized by patents more quickly than their corresponding journal articles. This enhances the visibility of early research results in patent processing, thereby strengthening the connection between science and technology. This acceleration is facilitated by the early access and open access nature of preprints, which enable faster innovation. Innovation in companies and some research institutes rely more heavily on preprints, while universities and hospitals still overwhelmingly prefer journal articles, emphasizing the role of peer-reviewed research in academic, clinical, and healthcare decision-making.

The dissemination speed of preprints in technological innovations appear to outpace that in scholarly communication, although it is not as widespread. Differences in the usage patterns between academic researchers and inventors are particularly evident in the post-online-publication citations of preprints by inventors, in contrast to academic researchers. The substantial number of post-online-publication patent citations highlights the critical role of the open science model, particularly the "open access" effect of preprints, in extending the influence of scientific research on industrial innovation and the economy. The patent citation advantage of bioRxiv-deposited articles remains evident even after accounting for factors related to publication venue and authorship.

The study enhances both the theoretical and empirical understanding of the role of preprints in advancing scientific and technological development and in building an open innovation ecosystem. The findings provide empirical evidence supporting the idea that open science policies, which encouraging the sharing of early research outputs like preprints, contribute to a more efficient connection between scientific research and technological innovations, thereby suggesting potential economic benefits and substantial returns on funding. Future research could explore the quality or economic value of patents that cite preprints by examining factors such as renewal status, citations, and transfer rate. Such analysis would offer a more comprehensive assessment of the impact of preprints on technological innovation. Additionally, it is crucial to distinguish whether citations originate from patent applicants or examiners, as their distinct motivations (Higham et al., 2021). Through such analyses, we can gain deeper insights into how preprints facilitate the transfer of scientific discoveries into technological innovations, thereby providing strong justification for optimizing open science policies and practices.

## Acknowledgements

The first author was supported by the Social Science Planning Foundation of Liaoning Province (grant L24CTQ002). The second author was supported by the Major Program of Chinese Ministry of Education (grant 22JZD021), National Natural Science Foundation of China (grant L2324108). Our gratitude also goes to the anonymous reviewers and the editor for their valuable comments.

## Appendix A